# Disjoint Paths Multi–stage Interconnection Networks Stability Problem

Ravi Rastogi[1], Nitin[1], Durg Singh Chauhan[2] and Mahesh Chandra Govil[3]

[1]Department of CSE and IT, Jaypee University of Information Technology,
Waknaghat, Solan, Himachal Pradesh 173234, India

[2]Uttarakhand Technical University,
Dehradun, Uttarakhand 241001, India

[3]Malaviya National Institute of Technology
Jaipur, Rajasthan 302017, India

**Abstract**
This research paper emphasizes that the Stable Matching problems are the same as the problems of stable configurations of Multi–stage Interconnection Networks (MIN). The authors have solved the Stability Problem of Existing Regular Gamma Multi-stage Interconnection Network (GMIN), 3-Disjoint Gamma Multi-stage Interconnection Network (3DGMIN) and 3-Disjoint Path Cyclic Gamma Multi-stage Interconnection Network (3DCGMIN) using the approaches and solutions provided by the Stable Matching Problem. Specifically Stable Marriage Problem is used as an example of Stable Matching. For MINs to prove Stable two existing algorithms are used:–the first algorithm generates the MINs Preferences List in $O(n^2)$ time and second algorithm produces a set of most Optimal Pairs of the Switching Elements (SEs) (derived from the MINs Preferences List) in $O(n)$ time. Moreover, the paper also solves the problem of Ties that occurs between the Optimal Pairs. The results are promising as the comparison of the MINs based on their stability shows that the ASEN, ABN, CLN, GMIN, 3DCGMIN are highly stable in comparison to HZTN, QTN, DGMIN. However, on comparing the irregular and regular MINs in totality upon their stability the regular MINs comes out to be more stable than the irregular MINs.

## 1. Introduction and Motivation

In a Stable Matching problem, the task is to match a number of persons in pairs, subject to certain preference information. Briefly, each person regards some of the others as acceptable mates and ranks them in order of preference. A matching is unstable if two persons did not match and considered together. The task is to find a stable matching, i.e., one that is acceptable to pairs. The subject of this paper is the Stable Marriage problem in particular, and stable matching problems in general. Gale and Shapley [1] first studied this problem. Gale and Shapley have shown that a stable matching always exists if the problem is a marriage problem, i.e., if the participants can be divided into two sexes, the men and the women, in such a way that the acceptable mates of each person are all of the opposite sex; in fact, both proposed a linear–time algorithm to find such a matching. Irving, in [3], gave a linear–time algorithm for general problem. An introductory treatment of stable matching appears in reference [4]; a comprehensive treatment is reported in reference [2].

This paper explores the relationship between stable matching and Multi–stage Interconnection Networks (MIN) Stability Problem. Mayr and Ashok proved that the Network Stability problem is NP–Complete in general [5–7], but when the network is a MIN, then the stability problem becomes equivalent to stable matching. Specifically, stable marriage problem has been used as an example of stable matching to solve the MINs stability problem. It is concluded that the situation in which the MINs become unstable and proved stable using the stable matching approach. The algorithms have been proposed to solve this problem and provide better solutions when the instances of stable matching have ties or when issues of deceit are involved. It explores the structure of all instances of stable matching similar to the research work presented by Gusfield [8], Irving [9] and Feder [10].

MINs are widely used for broadband switching technology and for multiprocessor systems. Besides this, MINs offers an enthusiastic way of implementing switches used in data communication networks. With the performance requirement of the switches exceeding several terabits/sec and teraflops/sec, it becomes imperative to make them dynamic and fault–tolerant [11–29]. In this paper, the stability problem of 8 x 8 Gamma Multi–stage

Interconnection Network (GMIN) [22, 23], 8 x 8 3–Disjoint Paths Gamma Multi–stage Interconnection Network (3DGMIN) [22, 23] and 8 x 8 3–Disjoint Paths Cyclic Gamma Interconnection Network (3DCGMIN) [22, 23] are solved and compared with the following Irregular MINs known as Hybrid ZETA Network (HZTN) [24], Quad–tree Network (QTN) [25] and Regular MINs known as Augmented Shuffle–exchange Network (ASEN) [26, 27], Augmented Baseline Network (ABN) [28] and Hybrid Cross-Link Network (CLN) [29]. The stability result of HZTN, QTN, ASEN, ABN and CLN are already evaluated and reported in [24].

The rest of the paper is organized as follows: Section 2 contains introduction of INs, MINs, and stable matching problems. Section 3 provides the algorithms, preference lists, and optimal pairs to solve the stability problems of 8 x 8 GMIN 8 x 8 3DGMIN and 8 x 8 3DCGMIN. Section 4 presents the results followed by the conclusion.

## 2 Preliminaries and Background
### 2.1 Stable Matching

An instance of stable matching is an instance of stable marriage if the persons divided into two sets, the men and the women, so that the acceptable mates of each person are all of the opposite sex. An instance of stable matching is an instance of Complete Stable Matching if there is an even number of persons and each person is acceptable to everyone else. Similarly, an instance of stable marriage is an instance of complete stable marriage if there are an equal number of men and women and each person is acceptable to every person of the opposite sex. The size of an instance of stable matching is the sum, over all persons x, of the number of persons acceptable to x. The most common tasks associated with an instance of stable matching are to determine whether a stable matching exists and to construct one if possible. Other tasks might include counting and enumerating all stable matchings of a given instance.

### 2.2 Multi–stage Interconnection Networks

MINs consist of multiple stages of SEs. Popular among them is a class of regular networks which in their basic form, consist of $\log_m N$ stages of m x m SEs connecting N input terminals to N output terminals. Sometimes MINs can also be built using large SEs and correspondingly have less number of stages, with similar properties. There exist many different topologies for MINs, which are characterized by the pattern of the between links between stages.

#### 2.2.1 Network Architecture of 8 x 8 Gamma Multi–stage Interconnection Network

A GMIN (see Figure 1) of size $N = 2^n$ has $n + 1$ stages labeled from 0 to $n$ and each stage involves $N$ switches. Switches of sizes 1 x 3 and 3 x 1 are coupled with the first and the last stage respectively. Each switch at intermediate stages is a 3 x 3 crossbar. Each switch $j$ at stage $I$ has three output link connections to switches at stage $(i + 1)$ according to the plus-minus-$2^i$ function. The $j^{th}$ switch at stage $I$ has three output links to switches $[(j - 2^i) \bmod N]$, $j$ and $[(j + 2^i) \bmod N]$ at each consecutive stage [1-5], [28-31].

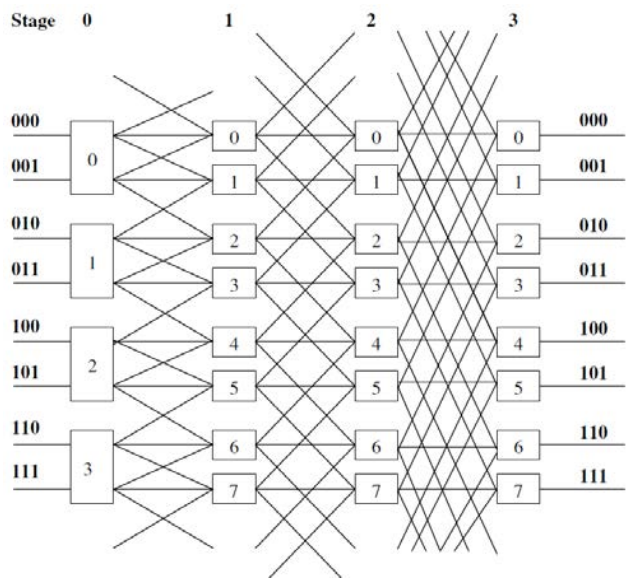

Fig. 1 A 8 x 8 Gamma Multi-stage Interconnection Network (GMIN).

#### 2.2.2 Network Architecture of 8 x 8 3–Disjoint Paths Gamma Multi–stage Interconnection Network

A 3DGMIN (see Figure 2) of size $N = 2^n$ is similar to gamma network, except the source nodes $2i$ and $2i + 1$ are combined into one 2 x 4 switch. These 2 x 4 switches deliver packets to
1. $i - 2, i - 1, i + 1$ and $i + 2$ (where $i$ is not equal to 0 or $N - 1$),
2. $i, i + 1, i + 2$ and $i + 3$ (where $i$ is equal to 0),
3. $i, i - 1, i - 2$ and $i - 3$ (where $i$ is equal to $N - 1$).

Similarly, the destination nodes $2i$ and $2i + 1$ are also combined into a 2 x 4 switch. These 2 x 4 switches recieve packets from
1. $i - 2, i - 1, i + 1$ and $i + 2$ (where $i$ is not equal to 0 or $N - 1$),
2. $i, i + 1, i + 2$ and $i + 3$ (where $i$ is equal to 0),
3. $i, i - 1, i - 2$ and $i - 3$ (where $i$ is equal to $N - 1$).

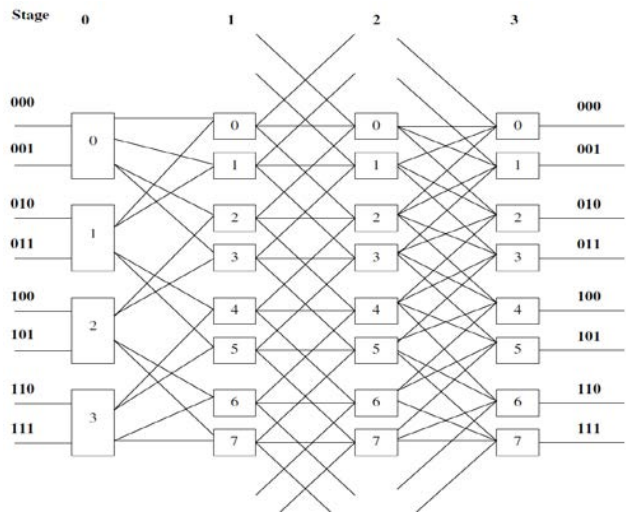

Fig. 2 A 8 x 8 3-Disjoint Gamma Multi-stage Interconnection Network (3DGMIN).

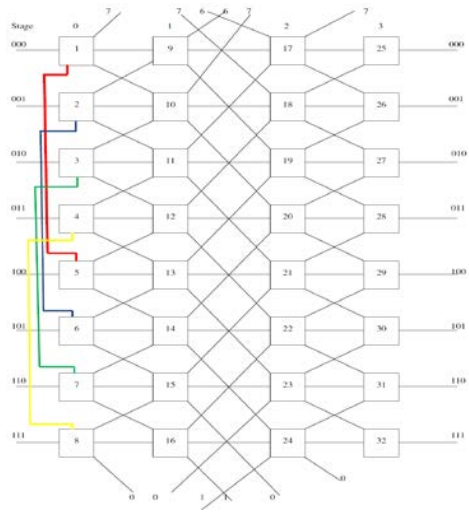

Fig. 3 A 8 x 8 3–Disjoint Paths Cyclic Gamma Interconnection Network (3DCGMIN).

### 2.2.3 Network Architecture of 8 x 8 3–Disjoint Paths Cyclic Gamma Interconnection Network

3DCGMIN (see Figure 3) is a cyclic Gamma Network, connecting $N = 2^n$ inputs to N outputs. It consists of $\log_2 N + 1$ stages with N switching elements per stage. The number of input nodes in 3DCGMIN are divided in two parts, and an alternate link is used connecting the respective inputs nodes to each other. It means that the first node in first part is connected with first node in second part with alternate link and so on. The 0th stage switches are 2 x 3 crossbars, 1st and 2nd stage switches are 3 x 3 crossbars, output stage switches are 3 x 1 crossbars. The connecting patterns between different stages are done as per CGMIN concept. The connections between stage $0 - 1$ & $2 - 3$ are done as per $2^0$ pattern whereas the $2^1$ pattern is used for connection between 1–2. Figure 3 shows the topology of 3DCGMIN for N = 8.

In 3DCGMIN, a packet visits n switches before reaching the destination. The stages are numbered 0 to n, from left to right. The connecting pattern between stages is given by plus–minus $2^{(\gamma+i) \bmod (n-1)}$ functions. The $j^{th}$ switch at stage i, $0 \le i \le n$, is connected with three switches at stage i+1 using three functions:

$$\text{fstraight}(j) = j$$
$$\text{fup}(j) = j - 2^{(\gamma+i) \bmod (n-1)} \bmod N$$
$$\text{fdown}(j) = j + 2^{(\gamma+i) \bmod (n-1)} \bmod N$$

The function fstraight defines the switch to be visited if a straight link is chosen. The functions fup and fdown denote the switches visited if we choose up and down links respectively. Each request in 3DCGMIN also carries a routing tag of n digits. Each digit in tag can take any of the following three values : 0, 1 and –1. We can use both the distance tag routing and destination tag routing methods to route a packet to its intended destination. By distance we mean Distance = D – S (Mod N), where D is the destination and S is the source. Following formula is used, to generate the all possible routing tags representing the distance between source and destination:

$$\text{RTDistance} = \delta_i^j 2^0 \pm \delta_i^j 2^1 \pm \delta_i^j 2^0$$

The alternate source / link at stage 0 is used in following cases: 1) The source S is faulty / non operational, 2) Source S is busy with packets and the current request needs urgent processing, 3) the buffer of source S is full, due to which the request is required to wait. The routing algorithm should make a decision about it. Whenever the packet is transferred to alternate source the routing needs one extra hop processing.

#### 2.2.3.1 Multiple Disjoint Paths

We can observe that the destination D at stage n is connected to three switches at stage n–1; $D + 2^\gamma$ (Mod N), D and $D - 2^\gamma$ (Mod N) respectively. Therefore, a path can reach to D through one of them. Therefore, the total number of alternative paths between source S and destination D should be sum of all possible paths from S to these three switches. We can estimate these numbers by using the recurrence relation used in CGMIN. If we consider the alternate source is also used for transmission then, the paths from it will prove additional to the original paths generated from S to D. It can be observed that, multiple paths are always present between every pair (S,

D), and we can get at least 3 disjoint paths considering the alternate link.

## 3. Solving Multi–stage Interconnection Networks Stability Problem using Stable Matching

3.1 MINs Stability Problem

MINs provide an easy way through which the information is routed via the specified switches, however it varies greatly with the type of topology that is used and it may be unstable for many instances. The routing mechanisms via the switches occur through the path–length algorithm upon the basis of which the shortest path to the destination is selected. Unstability in any MIN (regular or irregular) may occur if at any instance, a node fails and no alias path is available for routing through any of the nodes.

The switches are highly independent of each other as such no conjugation occurs amongst them thereby yielding no possible track and leaves the entire network as unstable. As the switches have no dependency, no backtracking mechanism is available thus if the initial nodes as in Figures (5, 11, and 15) fail the path is deadlocked and the entire network becomes unstable. The topology of the network has little significance associated with the unstability, as the network is not fault tolerant in case of failure thus unstability is bound to occur. The switches are unaware of the next immediate/most optimal path to follow to achieve successful delivery thereby deadlock remains causing unstability.

3.1.1 Conjugation

Here 3DCGMIN is taken as the example and the definition given here is remaining same for all other MIN discussed further. In Figure (3), both the subnetworks (i.e. $G^0$ and $G^1$) of CLN, have the conjugate pairs (in stage 0 of Figure (3), SE 1–16 forms a conjugate subset; within that subset, SE 1, SE 2, SE 3, SE 4, SE 5, SE 6, SE 7 and SE 8 are a conjugate pair; and SE 1 & 9, SE 2 & 10, SE 3 & 11, SE 4 & 12, SE 5 & 13, SE 6 & 14, SE 7 & SE 15 and SE 8 & SE 16 forms a conjugate loop).

3.2 Stable Matching and MINs Stability Problem

As mentioned in the above context due to the unstability of the network it becomes less fault–tolerant, which leads to deadlock situation. To tackle this problem a mechanism of stable matching is improvised to prevent failure from occurring. Since at every level n = 0, 1, 2, 3 the switches are aware of their immediate neighbors, chooses the best fit on the basis of the preference list created using the path–length approach from the preference matrix of the specific MIN.

With the application of the stable matching approach even in case of the path failure the conjugate pairs are active based on preference path length and the desired path is accepted. The same is carried forward for the backtracking approach thus no path is failed at every segment hence even in case of failures of first SE of each segment there is an alias path available so that the destination is reached. For example:
Case 1: Refer Figure (15). Now consider that you have send data from SE 1 to SE 25 and if SE 9 fails then the request either jumps to SE 10 (and follow the path = SE 1–SE 10–SE 18–SE 25) or it jumps to SE 5(and follow the path = SE1–SE5–SE12–SE18–SE25) using chaining link and reaches the destination using the path–length of 3 hence successfully transferring data that can be seen from the preference lists of the switches provided in Figure (13) resulting in keeping the entire network stable (as no congestion occurs).

3.3 Assumptions

Before writing the required algorithm, here are some assumptions that have to be taken care of. The assumptions while implementing the stable matching algorithm as following:

1. Conjugate Pairs of Nodes in the Network: The circuit thereby consists of segments $G^0$ and $G^1$ the corresponding alternate pairs of the levels 0, 1, 2 have conjugate roots between them thus the path can be traversed from these possible routes however it increases the net effective cost involved as it increases the specified path length.
2. Priority of the Traversal of the Paths: The algorithm that is employed in the calculation of paths is based on the concept of path length algorithm. Priority is given to the node by means of which the destination can be reached in minimum time and cost in comparison to any other node in the entire circuit.
3. Neglecting pairs of the level with minimum number of nodes: Since at the level with number of nodes the amount of inflowing paths is very high thereby the probability of it selecting the most optimized pair is very low, as it has multiple out flowing paths to the destination of relatively similar path lengths within the circuit and hence, get neglected.

The sole purpose of choosing the assumptions is the fact that without them the stability of the network cannot be proved. There are a large number of observable features present in the network that have to be neglected to prove the above cause. Assumptions have been included to enhance our effort to provide an efficient approach in proving the network to be stable. In addition, by assuming them, it helps us to decide the broad criteria of defining the constraints under which the network is going to act effectively and the concept of stable matching augmented well. The fundamental approach of assuming these conditions is to provide us with an initial approach that laid emphasis on the key aspect of stability using stable matching algorithm. Furthermore, if these assumptions are not considered then it will leads to NP–Completeness problem.

### 3.3.1 NP–Completeness

A problem is called NP (nondeterministic polynomial) if its solution (if one exists) can be guessed and verified in polynomial time, nondeterministic means that no particular rule is followed to make the guess. Thus, finding an efficient algorithm for any NP–Complete problem implies that an efficient algorithm can be found for all such problems, since any problem belonging to this class can be recast into any other member of the class. As far as the above solution is concerned the problem of NP–Completeness arises from the fact that the stability of the network can be rendered from the stable matching approach in a polynomial time solution hence the problem is solved. The solution of the optimal pairs of all the networks as per the algorithm is given above and is produced assuming into consideration of all the assumptions otherwise the solution fails.

Applying by the concept of stable matching, it will render us with an exact solution to the above problem in a defined polynomial time expression.

### 3.4 Algorithm for Deriving Preference Lists from the MINs

The algorithm to generate the preference lists of the MIN is explained here. This algorithm is on the similar lines of the Gale–Shapley Algorithm.

Algorithm: PREFERENCE_LISTS
___________________________________________
Inputs: Priority of SE/Nodes based on shortest path concept of reaching the goal.
Output: Provides a Priority Preference Lists from which the Optimal Pairs are selected.
Precondition: Each list has a collection of only those SE that in turn are always connected to.
Postcondition: The Optimized Preference lists are generated.
___________________________________________
1. Stable ← TRUE (No condition of Tie occurs with two SE having the same Priority Pairs)
2. FOR each Switch SE 1
3. FOR each Switch SE 2
4. IF ((SE 1 prefers SE 2 to its existing pair as it has a shorter path length to reach Destination SE and both are connected)) and
   ((SE 2 prefers SE 1 to its existing pair as it has a shorter path length to reach Destination SE and both are connected))
5. THEN the Switches SE 1 and SE 2 exist mutually in their list.
6. ELSE IF (If SE 1 and SE 2 have Tie for their list elements order them both in their lists)
7. ELSE (If SE 1 and SE 2 do not have a path amongst each other)
8. WRITE "SE 1 and SE 2 do not have a Stable Pair"
9. Stable ← FALSE
10. END IF
11. END IF
12. END FOR
13. END FOR
14. WRITE "The Preference Lists is generated"
15. EXIT
___________________________________________

Complexity: The run time complexity of the Algorithm: PREFERENCE_LISTS is $O(n^2)$.

Proof of Complexity or Correctness:
Let SE 1 = SE 2 = $n$
For lines from #2 to 13 the Time = $n \times n$ (time taken in generating the MINs preference lists) = $n^2 \times$ Constant

Therefore, Complexity in Big (O) notation is $O(n^2)$.

### 3.5 Reduction of the Ties in Irregular and Regular MINs

The reduction of the ties in the irregular and regular networks is discussed here. After deriving the preference lists of an irregular and regular network that has been created based on the patterns described in the previous section a basic aspect that has borne in mind is that while creating the preference list there are a large number of cases where ties occurs, which means for a specific SE that has to be resolved as it will result in congestion as two pairs have the same pair of optimal switches defined in the preference list. Thus in such a case priority is set in a such a way that the switch next in the list is tested for priority with all other switches and case of resolution of this clause it is allocated to the specific switch/ node and if this is not acceptable the procedure is carried on with other switches in the list and vice versa.

### 3.6 Deriving Optimal Pairs from MINs Preference Lists

The algorithm for a solution to a stable marriage instance in MIN is based on a sequence of "proposals" from one switch to the other based on shortest path length to reach the destination. Each switch proposes, in order, to the nodes (switches) on his preference list, pausing when a node agrees to consider his proposal, but continuing if a proposal is rejected either immediately or subsequently. When another node receives a proposal, it rejects it if the specified node already holds a better proposal, but otherwise agrees to hold it for consideration, simultaneously rejecting any poorer proposal that the node may currently hold i.e. the preference is given to the node which is higher or first in the priority list than any other nodes also specified later in its specific list.

It is not difficult to show, as in that the sequence of proposals so specified ends with every switch holding a unique proposal, and that the proposals held constitute a stable matching. The Two fundamental implications of this initial proposal sequence are:
1. If SE 1 pr oposes to SE 2, then there is no stable matching in which SE 1 has a better partner than SE 2.
2. If SE 2 receives a proposal from SE 1, then there is no stable matching in which SE 2 has a worse partner than SE 1.

These observations suggest us explicitly to remove SE 1 from SE 2's list and SE 2 from SE 1's list. If SE 1 receives a proposal from some node that is better in priority than SE 2 then the resulting lists or pairs as the shortlists for the given problem instance is referred.

Algorithm: SELECTING_STABLE_PAIRS

___
Inputs: Preference lists of SE.
Output: A matching consisting of list of engaged pairs.
Precondition: Each list includes the connection of one SE with all the other.
Postcondition: A matching is produced which is stable for each SE.
___
1. FOR each Switch SE
2. Engaged (SE) ← FALSE
3. END FOR
4. WHILE there is a SE which is not engaged
5. FOR each Switch SE y
6. IF Switch SE y is not yet engaged
7. THEN SE x ← highest on SE y list, which is not yet engaged
8. ADD (SE y, SE x) to the Stable Pair List
9. END IF
10. END FOR
11. END WHILE
12. Write "List of Optimal (Stable) Pairs"
___

Complexity: The run time complexity of the Algorithm: SELECTING_STABLE_PAIRS is $O(n)$.

Proof of Complexity or Correctness:

Let time of adding (SE y, SE x) to stable pair list = $t_1$

Number of SEs = $n$

Hence Time = $n \times t_1$

Therefore, Time Complexity in Big (O) notation is $O(n)$

## 3.7 Application of Stable Matching Approaches to Solve MINs Stability Problems

### 3.7.1 A 8 x 8 GMIN

It is know that the GMIN is regular networks and there is no need to give the path length algorithm, as the path length remains constant on all the routes (may be primary, secondary, or express).

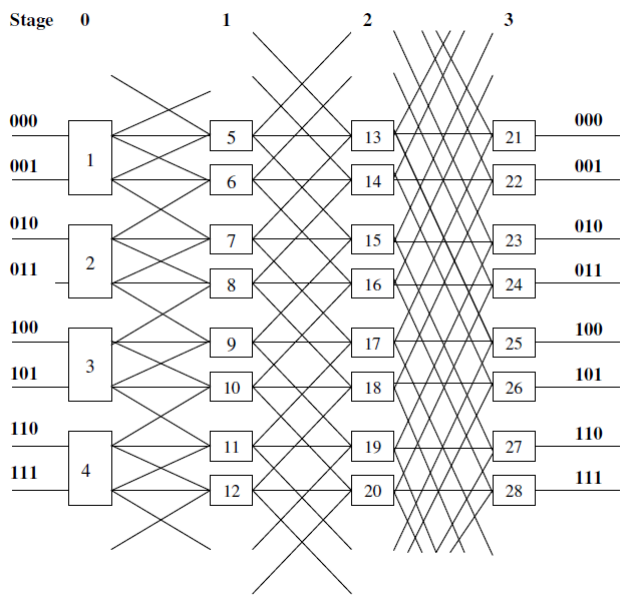

Fig. 4  A 8 x 8 GMIN. Here SEs are renumbered to solve the stability problem.

#### 3.7.1.1 Preference Lists

Refer algorithm explained in Section (3.4) for deriving preference lists for the GMIN. The SEs in Figure (1) are renumbered and put up again in Figure (4). Figure (5), shows the preference lists.

SE 1   5 6 7 13 15 14 16 15 13 17 21 25 23 27 22 26 24 28 25 21

SE 2   7 6 8 9 15 13 17 14 16 16 14 18 17 15 19 23 27 21 25 25 21 22 26 24 28 26 22 27 23

SE 3   9 8 10 11 17 15 19 16 14 18 18 16 20 19 17 25 21 23 27 27 23 24 28 22 26 26 22 28 24

SE 4   11 10 12 19 17 18 16 20 20 18 27 23 25 21 26 22 24 28 28 24

SE 5   13 15 21 25 23 27

SE 6   14 16 22 26 24 28

SE 7   15 13 17 23 27 21 25 25 21

SE 8   16 14 18 24 28 22 26 26 22

SE 9   17 15 19 25 21 23 27 27 23

SE 10  18 16 20 26 22 24 28 28 24

SE 11  19 17 27 23 25 21

SE 12  20 18 28 24 26 22

SE 13  21 25

SE 14  22 26

SE 15  23 27

SE 16  24 28

SE 17  25 21

SE 18  26 22

SE 19  27 23

SE 20  28 24

Fig. 5  The complete preference lists of the GMIN.

3.7.1.2 Reduction of the Ties

Refer procedure explained in Section (3.5). The same is used here to reduce the ties. See Figure (5) the preference list for the GMIN and from here we can conclude that the said network has no ties.

3.7.1.3 Deriving Optimal Pairs from the Preference Lists

Refer procedure explained in Section (3.6). The same is used here to derive the optimal pairs for the GMIN. The following Figure shows the optimal pairs for the GMIN.

(1, 5), (2, 7), (3, 9), (4,11),

(5,13), (6,14), (7,15), (8,16),

(9, 17), (10,18), (11,19), (12, 20),

(13,21), (14,22), (15,23), (16,24),

(17,25), (18,26), (19,27) and (20,28).

Fig. 6  The optimal pairs, which have been short–listed from the GMIN preference lists.

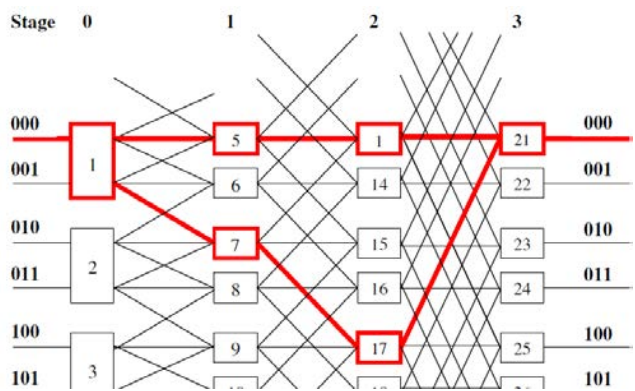

Fig. 7  The partial cut way part of GMIN.

Example 1. See Figure (7) and Table (1) for all possible routes and path–lengths. In this particular example a request is routed from source 0 to destination 0 i.e. source 0000 to destination 0000.

Table 1: The routing table of GMIN.

| Routes | Path–length |
|---|---|
| SE 1 – SE 5 – SE 13 – SE 21 | 3 |
| SE 1 – SE 7 – SE 17 – SE 21 | 3 |

Explanation: In this example (Table (4.4), all the possible paths from the source to destination are listed. To route a request from a given source to given destination can have possible routes and possible path–lengths. In the particular example, there are two paths from one source to destination. The first path (SE 1 – SE 5 – SE 13 – SE 21) is termed as the primary path, whereas the path (SE 1 – SE 7 – SE 17 – SE 21) is termed as the secondary path and will be used when the primary path is busy. The respective path–length at all the paths mentioned is 3 only. Since GMIN is a regular MIN, therefore it is always have a constant path–length on all the routes.

3.7.2 A 8 x 8 3DGMIN

It is know that the 3DGMIN is regular networks and there is no need to give the path length algorithm, as the path length remains constant on all the routes (may be primary, secondary, or express).

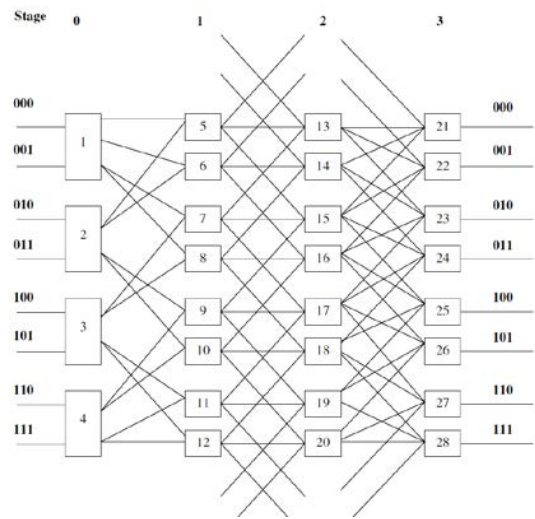

Fig. 8  A 8 x 8 3DGMIN. Here SEs are renumbered to solve the stability problem.

3.7.2.1 Preference Lists

Refer algorithm explained in Section (3.4) for deriving preference lists for the 3DGMIN. The SEs in Figure (2) are renumbered and put up again in Figure (8). Figure (9), shows the preference lists having Ties. All Ties have been solved for 3DGMIN.

```
SE 1    5 6 7 8 13 15 14 16 15 13 17 16 14 18 21 22 23 21 22 24 25 21 23 24 22 23 25 26 23 24 26 27 24 25 27 28
SE 2    5 6 9 10 13 15 14 16 17 15 19 18 16 20 21 22 23 21 22 24 25 21 23 24 22 23 25 26 23 24 26 27 25 26 24 25 27 28
SE 3    7 8 11 12 15 13 17 16 14 18 19 17 20 18 21 22 24 25 21 22 23 24 26 27 22 23 25 26 21 23 24 25 27 28 25 26 28 26 27 28
SE 4    9 10 11 12 17 15 19 18 16 20 19 17 20 18 23 24 26 27 21 22 24 25 25 26 28 24 25 27 28 22 23 25 26 27 28
SE 5    13 15 21 22 23 21 22 24 25
SE 6    14 16 21 23 24 22 23 25 26
SE 7    15 13 17 21 22 24 25 21 22 23 23 24 26 27
SE 8    16 14 18 22 23 25 26 21 23 24 24 25 27 28
SE 9    17 15 19 23 24 26 27 21 22 24 25 25 26 28
SE 10   18 16 20 24 25 27 28 22 23 25 26 26 27 28
SE 11   19 17 25 26 28 23 24 26 27
SE 12   20 18 26 27 28 24 25 27 28
SE 13   21 22 23
SE 14   21 23 24
SE 15   21 22 24 25
SE 16   22 23 25 26
SE 17   23 24 26 27
SE 18   24 25 27 28
SE 19   25 26 28
SE 20   28 26 27
```

Fig. 9 The complete preference lists of the 3DGMIN.

3.7.2.2 Reduction of the Ties

Refer procedure explained in Section (3.5). The same is used here to derive the optimal pairs for 3DGMIN. See Figure (9) the preference list for the SE 1 and SE 2 stands as:

SE 1    5 6 7 8 13 15 14 16 15 13 17 16 14 18 21 22 23 21 22 24 25 21 23 24

SE 2    5 6 9 10 13 15 14 16 17 15 19 18 16 20 21 22 23 21 22 24 25 21 23 24

Both the above cases have been rendered in such a method that SE 5 comes in priority 1 of them as such both can form the optimal pairs. Therefore to resolve the above conflict it is assumed that the SE 1 lays more emphasis upon considering the switch SE 5 first as it appears before hence it is allocated to it and for switch SE 2, SE 6 comes in the next order of preference and it is compared to all other members in the preference list in which SE 2 seems to have more priority over the switch SE 6 than any other switch hence is allocated to it. Thereby the optimal pairs are as follows:

SE 1 – – – SE 5
SE 2 – – – SE 6

The same procedure can be and is followed for all such cases in case such a collision occurs and a Tie for priority of switches occurs.

3.7.2.3 Deriving Optimal Pairs from the Preference Lists

Refer procedure explained in Section (3.6). The same is used here to derive the optimal pairs for the 3DGMIN. See Figure (9), the preference lists of 3DGMIN for the switch SE 5 stands as:

SE 5    13 15 21 22 23 21 22 24 25

SE 5 has highest priority been set to SE 13 as such appears first in the priority list and next priority has been set to SE 15 as such appears second and SE 21 as third and so on.

As it can be seen in Figure (9), that SE 7 has specified as:

SE 7    15 13 17 21 22 24 25 21 22 23 23 24 26 27

Thus the priority of SE 15 is more on the list of SE 7 thus both will exist as a stable matched pair and the set can be stable thus the above list reduced by eliminating the corresponding SE 15 from the list of SE 5 as someone else (SE 7) holds a better proposal for the SE 15 to follow the path and reach to its destination in minimum path length.
SE 5    13 – – – 21 22 23 21 22 24 25  and it becomes;
SE 5    13 21 22 23 21 22 24 25

Similarly, the nodes SE 15, SE 21, SE 22, SE 23, SE 24, SE 25 occur higher in the priority list of switches SE 7, SE 13, SE 16, SE 17, SE 18 and SE 19 thereby eliminating the above eight switches from the list of SE 5 and similarly the final list of optimal set is:

SE 5    13

Thus, the final pair becomes SE 5 and SE 13, which is stable in nature. Based on this assumption and analysis the following Figure (10) {which shows all the optimal pairs} has been compiled.

(1, 5),  (2, 6),  (3, 7),  (4, 9),

(5, 13), (6, 14), (7, 15), (8, 16),

(9, 17), (10,18), (11,19), (12, 20),

(13,21), (14,23), (15,22), (16,25),

(17,24), (18,27), (19,26) and (20,28).

Fig. 10 The optimal pairs, which have been short–listed from the 3DGMIN preference lists.

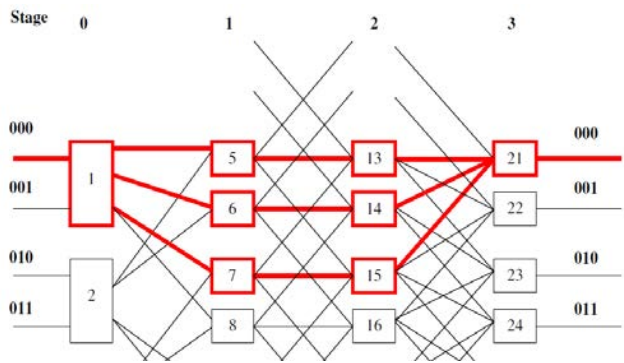

Fig. 11  The partial cut way part of 3DGMIN.

Example 2. See Figure (11) and Table (2) for all possible routes and path–lengths. In this particular example a request is routed from source 0 to destination 0 i.e. source 0000 to destination 0000.

Table 2: The routing table of 3DGMIN.

| Routes | Path–length |
|---|---|
| SE 1 – SE 5 – SE 13 – SE 21 | 3 |
| SE 1 – SE 6 – SE 14 – SE 21 | 3 |
| SE 1 – SE 7 – SE 15 – SE 21 | 3 |

Explanation: In this example (Table (2)), all the possible paths from the source to destination are listed. To route a request from a given source to given destination can have possible routes and possible path–lengths. In the particular example, there are three paths from one source to destination. The first path (SE 1 – SE 5 – SE 13 – SE 21) is termed as the primary path, the second path (SE 1 – SE 6 – SE 14 – SE 21) is termed as the secondary path whereas the path (SE 1 – SE 7 – SE 15 – SE 21) is termed as the third path and will be used when the primary and secondary paths are busy. The respective path–length at all the paths mentioned is 3 only. Since 3DGMIN is a regular MIN, therefore it is always have a constant path–length on all the routes.

3.7.3 A 8 x 8 3DCGMIN

It is known that the 3DCGMIN is a regular network and there is no need to give the path length algorithm, as the path length remains constant on all the routes (may be primary, secondary, or express).

3.7.3.1 Preference Lists

Refer algorithm explained in Section (3.4) for deriving preference lists for the 3DCGMIN. The SEs in Figure (3) are renumbered and put up again in Figure (12). Figure (15), shows the preference lists.

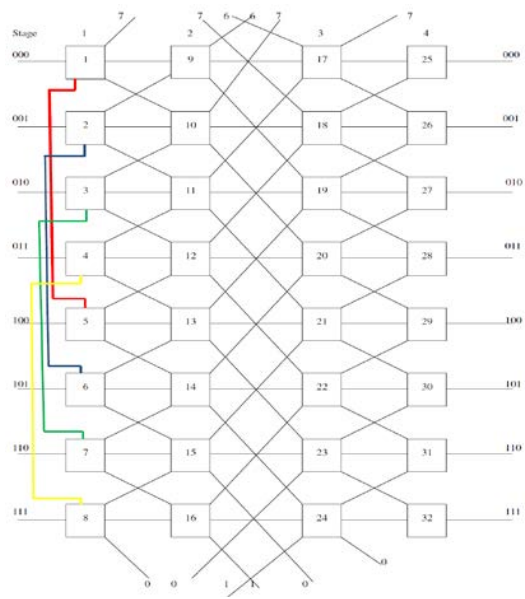

Fig. 12  A 8 x 8 3DCGMIN. Here SEs are renumbered to solve the stability problem.

SE 1   9 10 5 13 12 14 17 19 18 20 21 19 23 20 18 22 22 20 24 25 26 27 26 28 26 25 27 28 27 29 29 28 30 31 30 32 30 29 31 32 31

SE 2   10 9 11 6 14 13 15 18 20 17 19 19 17 21 22 20 24 21 19 23 23 21 26 25 27 28 27 29 25 26 27 26 28 29 28 30 29 31 32 31 31 32

SE 3   11 10 12 7 15 14 16 19 17 21 18 20 20 18 22 23 21 22 20 24 24 22 27 26 28 25 26 29 28 30 26 25 27 28 27 29 30 29 31 31 30 32 32 31

SE 4   12 11 13 8 1 6 1 5 20 18 22 19 17 21 21 19 23 24 22 23 21 28 27 29 26 25 27 30 29 31 27 26 28 25 26 29 28 30 31 30 32 32 31

SE 5   13 12 14 1 9 10 21 19 23 20 18 22 22 20 24 17 19 18 20 29 28 30 27 26 28 31 30 32 28 27 29 26 25 27 30 29 31 32 31 25 26

SE 6   14 13 15 2 10 9 11 22 20 24 21 19 23 23 21 18 20 17 19 19 17 21 30 29 31 28 27 29 32 31 29 28 30 27 26 28 31 30 32 26 25 27 25 26

SE 7   15 14 16 3 11 10 12 23 21 22 20 24 24 22 19 17 21 18 20 18 22 31 30 32 29 28 30 30 29 31 28 27 29 32 31 27 26 28 25 26 26 25 27

SE 8   16 15 4 12 11 13 24 23 23 21 20 18 22 19 17 21 21 19 23 32 31 30 32 29 28 30 28 27 29 26 25 27 30 29 31 27 26 28 25 26

SE 9   17 19 25 26 27 26 28

SE 10  18 20 26 25 27 28 27 29

SE 11  19 17 21 27 26 28 25 26 29 28 30

SE 12  20 18 22 28 27 29 26 25 27 30 29 31

SE 13  21 19 23 29 28 30 27 26 28 31 30 32

SE 14  22 20 24 30 29 31 28 27 29 32 31

SE 15  23 21 31 30 32 29 28 30

SE 16  24 22 32 31 30 29 21

SE 17  25 26

SE 18  26 25 27

SE 19  27 26 28

SE 20  28 27 29

SE 21  29 28 30

SE 22  30 29 31

SE 23  31 30 32

SE 24  32 31

Fig. 13  The complete preference lists of the 3DCGMIN.

#### 3.7.3.2. Reduction of the Ties

Refer procedure explained in Section (3.5). The same is used here to reduce the ties. See Figure (13) the preference list for the 3DCGMIN and from here, we can conclude that the said network has no ties.

#### 3.7.3.3 Deriving Optimal Pairs from the Preference Lists

Refer procedure explained in Section (3.6). The same is used here to derive the optimal pairs for the 3DCGMIN. The following Figure shows the optimal pairs for the 3DCGMIN.

(1,9), (2,10), (3,11), (4,12),

(5,13), (6,14), (7,15), (8,16),

(9, 17), (10,18), (11,19), (12, 20),

(13,21), (14,22), (15,23), (16,24),

(17,25), (18,26), (19,27), (20,28)

(21,29) and (22,30).

Fig. 14  The optimal pairs, which have been short–listed from the 3DCGMIN preference lists.

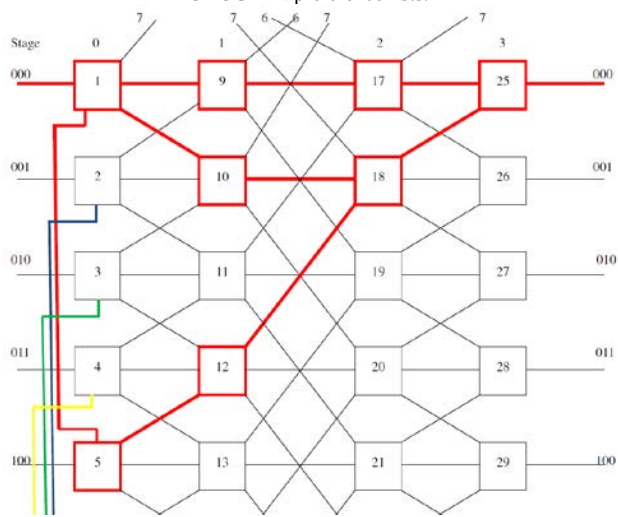

Fig. 15  The partial cut way part of ABN.

Example 3. See Figure (15) and Table (3) for all possible routes and path–lengths. In this particular example a request is routed from source 0 to destination 0 i.e. source 0000 to destination 0000.

Table 3: The routing table of 3DCGMIN.

| Routes | Path–length |
|---|---|
| SE 1 – SE 9 – SE 17 – SE 25 | 3 |
| SE 1 – SE 10 – SE 18 – SE 25 | 3 |
| SE 1 – SE 5 – SE 12 – SE 18 – SE 25 | 3 |

Explanation: In this example (Table (3)), all the possible paths from the source to destination are listed. To route a request from a given source to given destination can have possible routes and possible path–lengths. In the particular example, there are two paths from one source to destination. The first path (SE 1 – SE 9 – SE 17 – SE 25) is termed as the primary path, the second path (SE 1 – SE 10 – SE 18 – SE 25) is termed as the secondary path whereas the path (SE 1 – SE 5 – SE 12 – SE 18 – SE 25) is termed as the express path and will be used when the primary and the secondary paths are busy or faulty. The respective path–length at all the paths mentioned is 3 only. Since 3DCGMIN is a regular MIN, therefore it is always have a constant path–length on all the routes.

### 3.8 Comparisons

Based on the analysis of Sections (3.1–3.7) the comparison chart have been made and shown in Table (4) and Figure (16). It is depicted that the regular ASEN, ABN, CLN and 3DCGMIN are highly stable in comparison to the irregular HZTN, QTN and regular DGMIN as the neglected pairs (those who are not able to find any stable match) are 0 in their case. Therefore, regular MINs are highly stable according to the stable matching algorithm.

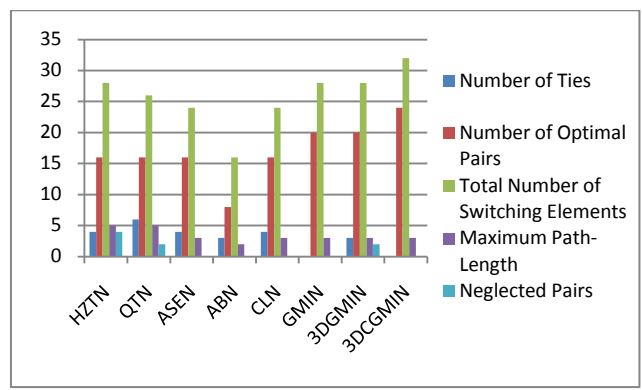

Fig. 16  The comparison graph of 16 x 16 different MINs based on their stability.

Table 4:  The comparison of 16 x 16 different MINs based on their stability.

| MINs | No. of Ties | No. of OPs/Total No. of SEs | Maximum PL | Neglected Pairs | MIN Status |
|---|---|---|---|---|---|
| HZTN | 4 | 16/28 | 5 | 4 | Low Stable |
| QTN | 6 | 16/26 | 5 | 2 | Intermediate Stable |
| ASEN | 4 | 16/24 | 3 | 0 | Highly Stable |
| ABN | 3 | 8/16 | 2 | 0 | Highly Stable |
| CLN | 4 | 16/24 | 3 | 0 | Highly Stable |
| GMIN | 0 | 20/28 | 3 | 0 | Highly Stable |

| | | | | | | |
|---|---|---|---|---|---|---|
| 3DGMIN | 3 | 20/28 | 3 | 2 | Intermediate Stable |
| 3DCGMIN | 0 | 24/32 | 3 | 0 | Highly Stable |

## 4. Conclusion

This paper explores the relationship between stable matching and MINs stability problem. Specifically stable marriage problem is used as example of stable matching to solve the MINs stability problem. The situations in which the fault–tolerant irregular and regular MINs become unstable have been shown. To counter this problem the appropriate algorithm, procedures, and methods have been designed using the concept of stable marriage. The ties problem of the optimal pairs has been solved. The comparison of the MINs based upon their stability shows that the ASEN, ABN, CLN, GMIN, 3DCGMIN are highly stable in comparison to HZTN, QTN and DGMIN. However, on comparing the irregular and regular MINs in totality upon their stability the regular MINs comes out to be more stable than the irregular MINs.

**Ravi Rastogi** is a PhD research scholar at Uttarakhand Technical University and currently working as a Senior Lecturer in the Department of Computer Science & Engineering and Information & Communication Technology, Jaypee University of Information Technology, Waknaghat, Solan-173234, Himachal Pradesh, India. He has published more than 10 research papers in the International refereed Journals and conferences.

**Nitin** is Ex. Distinguished Adjunct Professor of Computer Science, University of Nebraska, Omaha, USA. Currently he is working as an Associate Professor in the Department of Computer Science & Engineering and Information & Communication Technology, Jaypee University of Information Technology, Waknaghat, Solan, Himachal Pradesh, India. He has published more than 80 research papers in the reputed International refereed Journals and Conferences.

**Durg Singh Chauhan** is PhD from IIT Delhi and Postdoc from University of Maryland. Currently working as Vice Chancellor of Uttarakhand Technical University, Dehradun since 2009, Uttarakhand, India. He has published more than 125 r esearch papers in the reputed International refereed Journals and Conferences.

**Mahesh Chandra Govil** is M.Tech. and PhD from IIT Roorkee and Currently working as a Professor in the Department of Computer Science & Engineering at Malaviya Institute of Technology, Jaipur, Rajasthan, India. He has published more than 50 research papers in the refereed Journals and conferences.